\documentclass{article}
\usepackage[pdftex]{color,graphicx}
\usepackage{hyperref}
\begin{document}
\author{Kirana Kumara P\\Centre for Product Design and Manufacturing,\\Indian Institute of Science,\\Bangalore, Karnataka 560012, India\\\textit{email: kiranakumarap@gmail.com}
}
\title{Simulations using meshfree methods}
\maketitle
\begin{abstract}
In this paper, attempt is made to solve a few problems using the Polynomial Point Collocation Method (PPCM), the Radial Point Collocation Method (RPCM), Smoothed Particle Hydrodynamics (SPH), and the Finite Point Method (FPM). A few observations on the accuracy of these methods are recorded. All the simulations in this paper are three dimensional linear elastostatic simulations, without accounting for body forces.
\end{abstract}
\section{Introduction}

\subsection{Advantages of using meshfree methods}
(1) While using FEM to solve large deformation problems, considerable loss of accuracy may arise because of element distortions.

(2) Since it is not necessary to generate an element mesh while using meshfree methods, the methodology may be linked more easily with a CAD database than finite elements.

(3) No need of connectivity information

(4) Ideal for parallel processing

(5) Nodes can be added quite easily (h - adaptivity) in areas where more refinement is needed; thus accuracy may be easily controlled.

(6) Meshfree discretization can provide a representation of geometric object; the object representation is quite accurate; even the very complicated shapes may be represented easily. Geometry of an organ is represented by just a point cloud; there is no need of geometric entities or primitives. Accuracy of the object representation may be adjusted by increasing or decreasing the number of points used to represent the object.

(7) In FEM, cutting or tearing requires the computationally expensive remeshing process, and hence precomputed data of the object becomes invalid (at least locally) and all the data displayed to the user must be computed in realtime. But while using meshfree methods, only the equations corresponding to the nodes affected need to be modified.

(8) Just like FEM, pre-computations may be performed to improve efficiency.

(9) The property of lack of connectivity requirement coupled with their speed make some of the meshfree methods attractive for realtime simulations.

\subsection{Different meshfree methods}
Different meshfree methods differ by using (i) Different interpolating/approximating   functions (ii) Whether a strong form (which directly makes use of the governing equations) or a weak form (which makes use of an equivalent variational principle) or a combination of the two is used (iii) Type of the weighted residual method used (whenever applicable)

Some of the meshfree methods are: Smoothed Particle Hydrodynamics (SPH) method, Element Free Galerkin (EFG) method, Reproducing Kernel Particle Method (RKPM), Vortex method, Generalized Finite Difference method, Diffuse Element method, Meshless Local Petrov-Galerkin (MLPG) method, Partition of Unity method, h-p Cloud method, Particle-In-Cell method, Natural Element method, Method of Finite Spheres (MFS), Free Mesh method (FMM), Polynomial Point Collocation Method (PPCM), Radial Point Collocation Method (RPCM), Finite Point Method (FPM).

\subsection{Meshfree strong form methods}
Meshfree strong form methods (which directly make use of the governing equations) are faster but less accurate. But for some simulations, speed could be more important than accuracy (e.g., in the case of real-time simulation of biological organs, some amount of error may be allowed in the simulations, since humans cannot perceive small changes in the compliance). Hence a few meshfree strong form methods are tried out in the present paper with the intention of achieving fast simulations. The meshfree strong from methods that are tried out are: Polynomial Point Collocation Method (PPCM), Radial Point Collocation Method (RPCM), Smoothed Particle Hydrodynamics (SPH), and Finite Point Method (FPM). The next section deals with the use of PPCM, RPCM, and SPH, while the subsequent few sections deal with the use of FPM.

\section{Using PPCM, RPCM, and SPH}

A MATLAB code that can solve 3D linear elastostatics using PPCM was written from scratch. The code was used to solve problems where a cantilever beam was subjected to a prescribed end displacement or a prescribed end force. It was found that the code does not give accurate solutions; results (for a prescribed end displacement or for a prescribed end force both) were similar to the one shown in the Figure 1 (only qualitative, ignore the actual values).

\begin{figure}
\begin{center}
\includegraphics[width=0.5\textwidth]{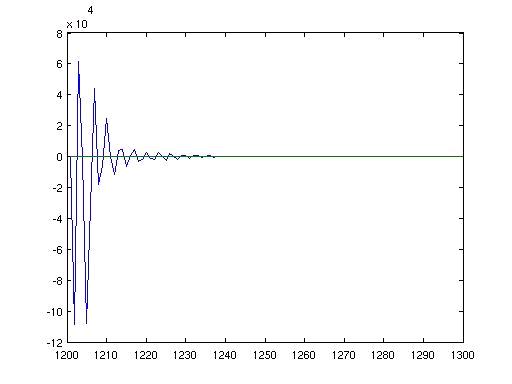}
\caption{Displacement of a cantilever beam subjected to an end force, using PPCM}          
\end{center} 
\end{figure}

Similarly, A MATLAB code that can solve 3D linear elastostatics using RPCM was written from scratch. The code was used to solve a problem where a cantilever beam was subjected to a prescribed end displacement. It was found that the code does not give accurate solutions; results (for a prescribed end displacement) was similar to the one shown in the Figure 2 (only qualitative, ignore the actual values).

\begin{figure}
\begin{center}
\includegraphics[width=0.5\textwidth]{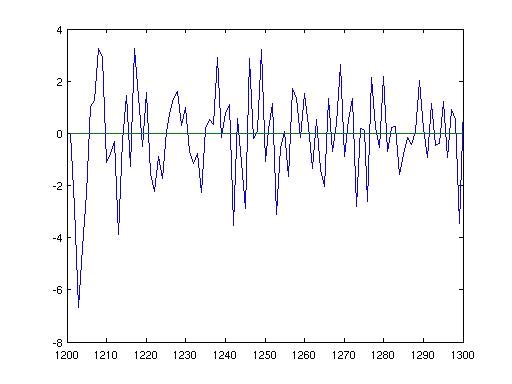}
\caption{Displacement of a cantilever beam subjected to an end displacement, using RPCM}          
\end{center} 
\end{figure}

Possible reasons for the failure of PPCM and RPCM: 

(1) They are not well developed, not well established.

(2) In the literature, PPCM and RPCM have been mostly used to solve 1D and 2D problems only; also, literature tells that results may be very bad when they are used to solve 3D problems.

(3) Both PPCM and RPCM use collocation technique; collocation introduces errors near boundaries since problem domain is of finite size (since the problem domain is discretized using only points). One can also note that the collocation technique itself is prone to instabilities.

Similarly, A MATLAB code that can solve 3D linear elastostatics using SPH was written from scratch. The code was used to solve problems where a cantilever beam was subjected to a prescribed end displacement or a prescribed end force. Again, it was found that the code does not give accurate solutions. Again, one can note that the literature tells that SPH may not give accurate results for problems where the domain is of finite size.

\section{Using the FPM}

Although FPM was successful in solving problems mentioned in a paper authored by this author [Kirana Kumara P, Ashitava Ghosal, 2012], upon using FPM to solve some other problems, it is found that FPM does not always give accurate solutions. For example, when a cantilever beam is subjected to an end force, the displacement (of the central axis) of the beam calculated using FPM is found to show instabilities, as shown in Figure 3. Also, the solution given by FPM is found to be incorrect here. While using FPM, one needs to choose values for some parameters but the values of the parameters are dependent on the problem at hand; and the values cannot be known beforehand. For the purpose of obtaining the solution shown in Figure 3, the values of the parameters are assumed to be the same as those used in the paper [Kirana Kumara P, Ashitava Ghosal, 2012]. Although it may be possible to choose the values of the parameters (by trial and error) in such a way that the solution obtained by using those values is the correct solution, one needs to know the correct solution even before the problem is solved using FPM in this case. But there is no need to use FPM if the solution is already known, and there is no use in using FPM if it cannot provide reliable solutions to problems whose solutions are not known beforehand.

The MATLAB code for FPM that is used to solve the above problem does not use stabilization techniques to stabilize the FPM. To see whether using stabilization techniques to stabilize FPM could improve the result for the problem related to Figure 3, the MATLAB code is improved to incorporate the stabilization technique explained in [E. Onate, F. Perazzo, J. Miquel, 2001]. But it is found that FPM still does not give accurate solution although instabilities (or oscillations in Figure 3) reduced.

The FPM used here uses the collocation technique. It is a well known fact that collocation techniques (that directly discretize governing equations) are inherently unstable. This may also be a reason for the instability and the inaccuracy associated with Figure 3.

\begin{figure}
\begin{center}
\includegraphics[width=0.5\textwidth]{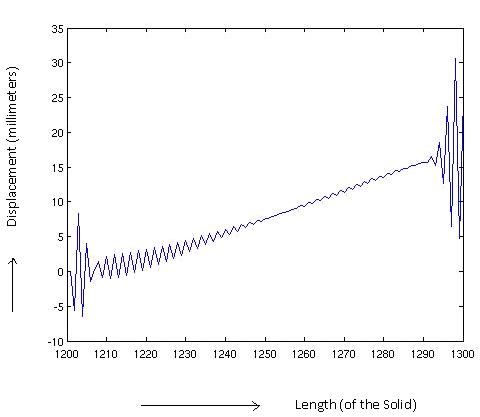}
\caption{Displacement along neutral axis for a cantilever with end force}          
\end{center} 
\end{figure}

Here, one should also note a point that is not duly noted in the literature. Meshfree methods like FPM use only points to discretize the problem domain. But a problem domain in 3D elastostatics is geometry, and hence the geometry is represented by just a set of points here. But it is well known that an object representation that is based on a point cloud is not unique in the sense that the conversion of the point cloud to a solid which is represented by its boundary (i.e., boundary representation) is not unique. Although stabilization techniques can take care of this problem to some extent (effectively they try to define the boundary of the problem domain in some way), they may not be providing a complete solution to the problem.

\section{Solving a 1D problem using FPM}

In the previous sections of this paper, only 3D problems are considered. The present section tries to solve a 1D linear elastostatic problem using FPM.

The problem is to find the displacement at one end of a bar where a force of 50 N is applied; the other end of the bar is fixed. The length of the bar is equal to 99 mm. The bar is discretized by equally spaced nodes, 100 nodes in total; hence, nodal spacing is equal to 1 mm. Area of cross section of the bar = 1 mm\textsuperscript{2}. Young's modulus = 200000 N/mm\textsuperscript{2}.

From analytical solution, tip displacement = 0.0248 mm.

Governing equation for the problem is given by
\[
E\frac{d^2u}{dx^2} = 0
\]

Now the solution obtained by using FPM is tabulated in Table 1, for different number of points in the support domain.

\begin{table}[c]
\begin{center}
\caption{Solution for a 1D problem for different number of points in the support domain}
\begin{tabular}{|c|c|}
\hline
\textbf{Number of points in the support domain} & \textbf{Solution by FPM (mm)} \\ \hline
3 & 0.0273 \\ \hline
5 & 0.0247 \\ \hline
10 & 0.0247 \\ \hline
15 & 0.0247 \\ \hline
20 & 0.0248 \\ \hline
30 & 0.0248 \\ \hline
40 & 0.0267 \\ \hline
50 & 0.0539 \\ \hline
60 & 0.1737 \\
\hline
\end{tabular}
\end{center}
\end{table}

By studying the table, one can conclude that the solution is accurate for a large range of the number of points in the support domain. But if the number of points in the support domain is either too low or too high, the solution tends to be inaccurate. The reason for the inaccuracy is that when there are too few points in the support domain, condition number of the matrix used to obtain \boldmath$\alpha$ (as explained in [E. Onate, F. Perazzo, J. Miquel, 2001]) becomes very small and this results in inaccurate results; and when there are too many points in the support domain, the condition number of the global `stiffness' matrix becomes very small and this is responsible for the inaccurate results.

\section{Summary}

Present paper has tried to solve some problems by using PPCM, RPCM, and SPH, but the results obtained were inaccurate. Possible reasons for the failure have been noted down; the reasons are deduced based partly on the available literature and partly on the present author's own insights.

Next, it was also found that FPM also cannot solve all the problems accurately. Reasons for getting inaccurate results could be because it is possible to get different solutions for the same problem if (i) number of points in the support domain is altered (ii) different values are chosen for certain parameters. Adding to these problems, the collocation method used in FPM is inherently unstable. Further, the object representation used in FPM is not fully unambiguous. These may be the reasons why accurate solutions could not be obtained for some of the problems considered in the present paper, even after incorporating a stabilization technique. Some of these observations are not found anywhere in the literature, and the present author has himself figured out some of the above reasons for the inaccurate results given by FPM in some cases.

In conclusion, FPM is not guaranteed to give accurate solutions. Whenever a solution given by FPM is found to be inaccurate, it may be possible to adjust the parameters, to adjust the number of points in the support domain, and to use a stabilization technique, such that FPM would give accurate results after making the adjustments; but one needs to know the correct solution beforehand to achieve this. But one can see that there is no need to obtain a solution to a problem if the correct solution to the problem is already known.

Many other meshfree techniques (which are considered fast) have some similarities (like the possibility of choosing different number of points in the support domain, possibility of using different values for certain parameters, ambiguity in the object representation (especially the boundary of an object) because of representing a solid by just a set of points etc.) with FPM, and hence they are likely to suffer the same drawbacks as FPM when it comes to the issue of accuracy. Further, many of the meshfree methods are not well developed, and some methods have not been used to solve 3D problems yet; literature on some methods is scant, and one comes to know the drawbacks of the methods only after writing codes from scratch and applying the method to solve some real-world problems. Of course, some of the meshfree methods are well established (e.g., RKPM) but the literature tells that they are slow.

Codes for solving elasticity problems using meshfree methods are not readily available, and hence the present author had to write many codes from scratch. Present author has written MATLAB codes (from scratch) that can solve 3D linear elastostatic problems using PPCM, RPCM, SPH, and FPM. But only FPM codes are provided as two appendices. Appendix I provides a MATLAB code for one dimensional linear elastostatics using the Finite Point Method, while Appendix II provides a MATLAB code for three dimensional linear elastostatics using the Finite Point Method.

\section{Acknowledgement}
Author acknowledges the Robotics Lab, Department of Mechanical Engineering, Indian Institute of Science, Bangalore, INDIA, for providing the necessary infrastructure to carry out this work.

\section{References}
Kirana Kumara P, Ashitava Ghosal, 2012, Real-time Computer Simulation of Three Dimensional Elastostatics using the Finite Point Method, Applied Mechanics and Materials, 110-16, pp. 2740-2745

E. Onate, F.Perazzo, J. Miquel, 2001, A finite point method for elasticity problems, Computers and Structures, 79 (2001), pp. 2151- 2163

\section{Appendix I: A MATLAB code for one dimensional linear elastostatics using the Finite Point  Method}

The code solves a one dimensional problem where one end of the bar is fixed while the other end is  subjected to a force of 50 N. Length of the bar = 99 mm, nodal spacing = 1 mm, total number of  nodes = 100 nodes, Young's modulus = 200000 N/mm\textsuperscript{2}, area of cross section = 1  mm\textsuperscript{2}.

\begin{verbatim}

1     clear;
2     clear all;
3     clc;
4     E=2*10^5;
5     x0=linspace(0,99,100);
6     n=20;
7     l=1;
8     for i=1:100           
9         gnode(l)=[x0(i)];
10        l=l+1;        
11    end
12    K=zeros((l-1),(l-1));
13    F=zeros((l-1),1);
14    for m=1:(l-1)
15        for v=1:(l-1)
16            dist(v)=((gnode(v)-gnode(m))^2)^(0.5);
17        end
18        [distn,NODE]=sort(dist);
19        rmax=distn(n);
20        rm=2*rmax;
21        c=0.25*rmax;
22        weights=(exp((-(distn.^2))/c^2)-exp((-rm^2)/c^2))/(1-exp((-rm^2)/c^2));
23        B=[];
24        A=zeros(3,3);
25        for o=1:n
26            x=gnode(NODE(o));
27            P=[1 x x^2]';
28            A=A+weights(o)*(P*P');
29            %A=A+(P*P');
30            B(:,o)=weights(o)*P;
31            %B(:,o)=P;
32        end
33        dhox2P=[0 0 2];    
34        dhox2N=dhox2P*(inv(A)*B);    
35        for q=1:n
36            K(m,NODE(q))=E*dhox2N(1,q);
37        end
38    end    
39    K(1,:)=0;
40    K(1,1)=1;    
41    for m=100:100
42        for v=1:(l-1)
43            dist(v)=((gnode(v)-gnode(m))^2)^(0.5);
44        end
45        [distn,NODE]=sort(dist);
46        rmax=distn(n);
47        rm=2*rmax;
48        c=0.25*rmax;
49        weights=(exp((-(distn.^2))/c^2)-exp((-rm^2)/c^2))/(1-exp((-rm^2)/c^2));
50        B=[];
51        A=zeros(3,3);
52        for o=1:n
53            x=gnode(NODE(o));               
54            P=[1 x x^2]';
55            A=A+weights(o)*(P*P');
56            %A=A+(P*P');
57            B(:,o)=weights(o)*P;
58            %B(:,o)=P;
59        end    
60        dhoxP=[0 1 2*x];
61        dhoxN=dhoxP*(inv(A)*B);    
62        for q=1:n        
63            K(m,NODE(q))=K(m,NODE(q))*(-0.5)*rmax;        
64            K(m,NODE(q))=K(m,NODE(q))+E*dhoxN(1,q);        
65        end
66        F(m)=50;
67    end
68    U=K\F;
69    U(100)
70    plot(x0,U)

\end{verbatim} 

\section{Appendix II: A MATLAB code for three dimensional linear elastostatics using the Finite Point  Method}

The code solves a three dimensional problem where a beam of length = 99 mm and 4 mm by 4 mm  cross section is discretized by uniformly spaced nodes located 1 mm apart. Young's modulus =  200000 N/mm\textsuperscript{2}, Poisson's ratio = 0.33 One end of the beam is completely fixed  (i.e., all the 25 nodes located on this end are fixed in all the three x, y, and z directions) while the mid -node located on the other end of the beam is subjected to a displacement of 5 mm in the y-direction  only. 

Author has written many codes very similar to the below code but with different boundary  conditions (like both ends of the beam fixed while a prescribed displacement is applied at the center of  the beam, one end of the beam fixed while a prescribed force is applied at the other end of the beam  etc.), using a different order of the polynomial (e.g., cubic and quintic polynomials instead of the  quadratic polynomial that is used in the below code), using lesser or more number of points in the  support domain, using different weight functions, and with or without using the stabilization technique.  The many similar codes written by the author also differ by the way the stabilization technique is  applied to the points located on edges or corners of the geometry (e.g., considering duplicate nodes at  the edges or corners, replacing edge by chamfer, resolving the normal at the nodes located on  edges/corners, refining discretization, splitting the nodes located on edges or corners into multiple  nodes and separating the multiple nodes by negligibly small distance, splitting the nodes located on  edges or corners into multiple nodes and separating the multiple nodes by negligibly small distance by  making use of random numbers, using a discretization which is not homogeneous and using a fine  discretization near the edges or corners etc.). All these codes are not provided here, since that would  need many pages and also since those codes are just a modification of the below code to incorporate  the necessary changes.

\begin{verbatim}

1     clear;
2     clear all;
3     clc;
4     E=2*10^5;
5     poisson=0.33;
6     lambda=(poisson*E)/((1+poisson)*(1-2*poisson));
7     mu=E/(2*(1+poisson));
8     x0=linspace(1,5,5);
9     y0=linspace(1,5,5);
10    z0=linspace(1,100,100);
11    n=50;
12    l=1;
13    for i=1:5
14    for j=1:5
15    for k=1:100
16    gnode(l,:)=[x0(i) y0(j) z0(k)];
17    l=l+1;
18    end
19    end
20    end
21    K=zeros(3*(l-1),3*(l-1));
22    F=zeros(3*(l-1),1);
23    for m=1:(l-1)
24    for v=1:(l-1)
25    dist(v)=(sum((gnode(v,:)-gnode(m,:)).^2))^(0.5);
26    end
27    [distn,NODE]=sort(dist);
28    rmax=distn(50);
29    rm=2*rmax;
30    c=0.25*rmax;
31    weights=(exp((-(distn.^2))/c^2)-exp((-rm^2)/c^2))/(1-exp((-rm^2)/c^2));
32    B=[];
33    A=zeros(10,10);
34    for o=1:n
35    xyz=gnode(NODE(o),:);    
36    x=xyz(1,1);
37    y=xyz(1,2);
38    z=xyz(1,3);
39    P=[1 x y z x^2 y^2 z^2 x*y y*z x*z]';
40    A=A+weights(o)*(P*P');
41    %A=A+(P*P');
42    B(:,o)=weights(o)*P;
43    %B(:,o)=P;
44    end
45    dhox2P=[0 0 0 0 2 0 0 0 0 0];
46    dhoy2P=[0 0 0 0 0 2 0 0 0 0];
47    dhoz2P=[0 0 0 0 0 0 2 0 0 0];
48    dhoxdhoyP=[0 0 0 0 0 0 0 1 0 0];
49    dhoydhozP=[0 0 0 0 0 0 0 0 1 0];
50    dhoxdhozP=[0 0 0 0 0 0 0 0 0 1];
51    dhox2N=dhox2P*(inv(A)*B);
52    dhoy2N=dhoy2P*(inv(A)*B);
53    dhoz2N=dhoz2P*(inv(A)*B);
54    dhoxdhoyN=dhoxdhoyP*(inv(A)*B);
55    dhoydhozN=dhoydhozP*(inv(A)*B);
56    dhoxdhozN=dhoxdhozP*(inv(A)*B);
57    for q=1:n
58    K((3*m-2),(3*NODE(q)-2))=K((3*m-2),(3*NODE(q)-2))+(lambda+mu)*dhox2N(1,q) 
                                 +mu*dhox2N(1,q)+mu*dhoy2N(1,q)+mu*dhoz2N(1,q);
59    K((3*m-2),(3*NODE(q)-1))=K((3*m-2),(3*NODE(q)-1))+(lambda+mu)*dhoxdhoyN(1,q);
60    K((3*m-2),3*NODE(q))=K((3*m-2),3*NODE(q))+(lambda+mu)*dhoxdhozN(1,q);
61    K((3*m-1),(3*NODE(q)-2))=K((3*m-1),(3*NODE(q)-2))+(lambda+mu)*dhoxdhoyN(1,q);
62    K((3*m-1),(3*NODE(q)-1))=K((3*m-1),(3*NODE(q)-1))+(lambda+mu)*dhoy2N(1,q) 
                                 +mu*dhox2N(1,q)+mu*dhoy2N(1,q)+mu*dhoz2N(1,q);
63    K((3*m-1),3*NODE(q))=K((3*m-1),3*NODE(q))+(lambda+mu)*dhoydhozN(1,q);
64    K(3*m,(3*NODE(q)-2))=K(3*m,(3*NODE(q)-2))+(lambda+mu)*dhoxdhozN(1,q);
65    K(3*m,(3*NODE(q)-1))=K(3*m,(3*NODE(q)-1))+(lambda+mu)*dhoydhozN(1,q);
66    K(3*m,3*NODE(q))=K(3*m,3*NODE(q))+(lambda+mu)*dhoz2N(1,q)+mu*dhox2N(1,q) 
                                       +mu*dhoy2N(1,q)+mu*dhoz2N(1,q);
67    end
68    end
69    K(3899,3899)=1;
70    K(3899,1:3898)=0;
71    K(3899,3900:(3*(l-1)))=0;
72    F(3899)=5;
73    for r=0:24
74    gfixed(r+1)=100*r+1;
75    end
76    for s=1:25
77    K((3*gfixed(s)-2),(3*gfixed(s)-2))=1;
78    K((3*gfixed(s)-2),1:(3*gfixed(s)-3))=0;
79    K((3*gfixed(s)-2),(3*gfixed(s)-1):(3*(l-1)))=0;
80    K((3*gfixed(s)-1),(3*gfixed(s)-1))=1;
81    K((3*gfixed(s)-1),1:(3*gfixed(s)-2))=0;
82    K((3*gfixed(s)-1),(3*gfixed(s)):(3*(l-1)))=0;
83    K((3*gfixed(s)),(3*gfixed(s)))=1;
84    K((3*gfixed(s)),1:(3*gfixed(s)-1))=0;
85    K((3*gfixed(s)),(3*gfixed(s)+1):(3*(l-1)))=0;
86    end
87    for s=1:25
88    F(3*gfixed(s)-2)=0;
89    F(3*gfixed(s)-1)=0;
90    F(3*gfixed(s))=0;
91    end
92    U=K\F;
93    z1=linspace(1201,1300,100);
94    z=1;
95    for z2=1201:1300
96        z3(z)=U(3*z2-1);
97        z=z+1;
98    end
99    plot(z1,z3)

\end{verbatim}

\end{document}